\documentclass[final]{aipproc}
\layoutstyle{6x9}
\usepackage{graphicx,amsmath,wrapfig,epsfig}
\begin{document}
\title{Nuclear effects in neutrino-nucleus DIS}
\author{M. Hirai}{
address={Department of Physics, Faculty of Science and Technology,
Tokyo University of Science \\ 
2641, Yamazaki, Noda, Chiba, 278-8510, Japan}}
\author{S. Kumano}{
  address={KEK Theory Center, Institute of Particle and Nuclear Studies, KEK \\
           and 
           Department of Particle and Nuclear Studies,
           Graduate University for Advanced Studies \\
           1-1, Ooho, Tsukuba, Ibaraki, 305-0801, Japan}}
\author{K. Saito}{
address={Department of Physics, Faculty of Science and Technology,
Tokyo University of Science \\ 
2641, Yamazaki, Noda, Chiba, 278-8510, Japan}}

\begin{abstract}
We explain the current status of nuclear parton distribution functions 
in connection with neutrino-nucleus interactions.
Neutrino deep inelastic scattering (DIS) measurements have been done
for heavy nuclear targets such as iron and lead. In order to extract
structure functions of the nucleon, one needs to remove nuclear effects
from the data. However, recent studies indicate that there are
inconsistencies in nuclear modifications between charged-lepton
and neutrino scattering measurements. 
Nuclear medium effects could be also an origin 
for the NuTeV anomaly in the weak-mixing angle. In addition,
the modifications could affect neutrino-oscillation experiments 
because some DIS events of neutrino-oxygen nucleus interactions
are contained. On the other hand, the nuclear medium effects themselves
are interesting and important for describing nuclei in terms of
quark and gluon degrees of freedom.
\end{abstract}
\maketitle

\vspace{-0.8cm}
%%%%%%%%%%%%%%%%%%%%%%%%%%%%%%%%%%%%%%%%%%%%%%%%%%%%%%%%%%%%%%%%%%%%%%%%%%%%%%%%
%%%%%%%%%%%%%%%%%%%%%%%%%%%%%%%%%%%%%%%%%%%%%%%%%%%%%%%%%%%%%%%%%%%%%%%%%%%%%%%%
\section{Introduction}
\vspace{-0.3cm}

Nuclear modifications of structure functions were found 
in charged-lepton deep inelastic scattering (DIS) by
the European Muon Collaboration (EMC) \cite{emc88},
so that they are often called the EMC effect.
Since nuclear binding energies per nucleon are much smaller than
a typical DIS scale, such effects were not expected except
for possible Fermi-motion effects. Later, the nuclear effects are
interpreted mainly by binding and Fermi-motion mechanisms with
possible internal modifications of the nucleon at medium and
large $x$ \cite{st94}.
At small $x$, suppression is observed in the structure functions,
and the phenomenon is called shadowing which is
understood by multiple scattering of $q \bar q$-like states
originating from a virtual photon. The reader may look at
a summary article in Ref. \cite{sumemc} for explanations
of these mechanisms.

On the other hand, there are growing interests in nuclear effects
in neutrino DIS structure functions with the following reasons.
\begin{itemize}
\vspace{-0.2cm}
\item[1.]
Current neutrino DIS measurements contain nuclear medium 
effects because heavy nuclear targets, iron and lead,
are used. Neutrino data are valuable especially for determining
valence-quark distributions in the nucleon because the parity-violating
$F_3$ structure function directly probes the valence distributions.
However, the nuclear effects should be properly removed from
the neutrino measurements for extracting information on 
the nucleonic parton distribution functions (PDFs) \cite{mstw09}.
\vspace{-0.0cm}
\item[2.]
There is an issue of NuTeV anomaly on the weak-mixing angle 
$\sin^2 \theta_W$ \cite{nutev-anomaly}. It could be related
to a nuclear medium effect in the iron nucleus \cite{sinth}.
\vspace{-0.0cm}
\item[3.]
For neutrino oscillation measurements, nuclear effects need
to be understood for neutrino-oxygen nucleus reactions \cite{sakuda}. 
Although such experiments are not at very
high energies, they contain a certain number of DIS events. 
In order to understand the DIS as well as a resonance region,
quark-hadron duality should be studied for extending the knowledge
of nuclear DIS to a relatively low-energy region \cite{duality}.
\end{itemize}
\vspace{-0.15cm}

In order to determine nuclear effects in general, a global $\chi^2$
analysis method was developed in Ref. \cite{hkm01} by using
all the available charged-lepton DIS data, and then by
adding Drell-Yan data to the data set \cite{hkn04}. 
Subsequently, there have been various proposals for 
the optimum nuclear PDFs. Here, we explain recent progress
on the various global analyses
\cite{ds04,hkn07,sykmoo08,eps09,kp07,fgs05}.
Nuclear parton distributions have been determined in Refs. 
\cite{ds04,hkn07,sykmoo08,eps09}, whereas Ref. \cite{kp07}
focused on structure functions by using conventional nuclear models
and Ref. \cite{fgs05} especially on the shadowing part.

\vspace{-0.2cm}
%%%%%%%%%%%%%%%%%%%%%%%%%%%%%%%%%%%%%%%%%%%%%%%%%%%%%%%%%%%%%%%%%%%%%%%%%%%%%%%%
%%%%%%%%%%%%%%%%%%%%%%%%%%%%%%%%%%%%%%%%%%%%%%%%%%%%%%%%%%%%%%%%%%%%%%%%%%%%%%%%
\section{Structure functions in neutrino DIS}
\vspace{-0.2cm}

Neutrino charged-current interactions with the nucleon are
described by the matrix element
\vspace{-0.35cm}
\begin{equation}
M = \frac{{G_F /\sqrt 2 }}{{1 + Q^2 /M_W^{{\rm{ }}2} }}
{\rm{ }}\bar u(k',\lambda '){\rm{ }}\gamma ^\mu (1 - \gamma _5 {\rm{) }}
u(k,\lambda ){\rm{  }}\left\langle X \right|{\rm{ }}J_\mu ^{CC} (0)
\left| {p,\lambda _N } \right\rangle ,
\end{equation}
where $G_F$ is the Fermi coupling constant,
$M_W$ is the $W$ mass, 
$Q^2$ is given by $Q^2=-q^2$ with the four-momentum transfer $q$,
$k$ ($\lambda$) and $k'$ ($\lambda'$) indicate initial and final lepton
momenta (spins), $p$ ($\lambda_N$) is the nucleon momentum (spin),
and $J_\mu ^{CC} (0)$ is the weak charged current (CC) of the nucleon.
The absolute-value square $|M|^2$ is calculated with an average
over the nucleon spin for obtaining the unpolarized cross section.
The leptonic part is calculated and the hadronic part
becomes the hadron tensor $W_{\mu\nu}^{CC}$, which is then expressed
by three structure functions $F_1$, $F_2$, and $F_3$:
\begin{equation}
\! 
\left( {\frac{{d\sigma }}{{dxdy}}} \right)_{CC}^{\nu ,\bar \nu } 
\! \! \!
= \frac{{G_F^2 M_N E}}{\pi{(1 + Q^2 /M_W^2 )^2 }}
\left[ {F_1^{cc} xy^2  + F_2^{cc}
\left( {1 - y - \frac{{M_N xy}}{{2E}}} \right) 
\pm F_3^{cc} xy\left( {1 - \frac{y}{2}} \right)} \right] ,
\! \!
\end{equation}
where $\pm$ indicates $+$ for $\nu$ and $-$ for $\bar \nu$,
$s$ is the center-of-mass energy squared,
$x$ is the Bjorken scaling variable $x=Q^2/(2 p\cdot q)$,
$y$ is defined by $y=p\cdot q/(p\cdot k)$,
$E$ is the neutrino-beam energy,
and $M_N$ is the nucleon mass.
The cross section is calculated in a quark-parton model to 
express the structure functions in terms of parton distribution functions
(PDFs):
\begin{alignat}{3}
              F_2  & = 2xF_1, \ \ \ \ \  &
      F_2^{\nu p}  & = 2x(d + s + \bar u + \bar c), \ \ \ \ \  &
 F_2^{\bar \nu p}  & = 2x(u + c + \bar d + \bar s),     
\nonumber   \\ 
                \  &  \  &
      xF_3^{\nu p} & = 2x(d + s - \bar u - \bar c), \ \ \ \ \  &
 xF_3^{\bar \nu p} & = 2x(u + c - \bar d - \bar s) ,     
\label{eqn:f23}
\end{alignat}
in the leading twist and the leading order of the running
coupling constant $\alpha_s$. These equations are combined to become
\begin{equation}
F_3^{\nu p}  + F_3^{\bar \nu p}  
= 2(u_v  + d_v ) + 2(s - \bar s) + 2(c - \bar c) ,
\end{equation}
for the structure functions $F_3$. It indicates that
measurements of $F_3$ are valuable for determining
the valence-quark distributions in the nucleon
because $s-\bar s$ and $c-\bar c$ are considered to be very small.
Another important point of neutrino DIS is to 
determine the strange quark ratio to the light-quark ones
$2 \bar s /(\bar u+\bar d)$ by using neutrino-induced
opposite-sign dimuon events
($ \nu _\mu  p \to \mu ^ +  \mu ^ -  X $)
by assuming a charm-quark production process.

%%%%%%%%%%%%%%%%%%%%%%%%%%%%%%%% table %%%%%%%%%%%%%%%%%%%%%%%%%%%%%%%%%%%%%%
\begin{wraptable}{r}[0.0cm]{0.52\textwidth}
{\small \hspace{0.3cm} {\bf TABLE 1.}  Recent neutrino DIS experiments}
\label{tab:experiments}
       \begin{minipage}[c]{0.3cm}
       \ \ 
       \end{minipage}
\small
\begin{center}
\begin{tabular}{lcc}
\hline
Experiment	& Target & $\nu$ energy (GeV) \\
\hline
CCFR        & Fe     & 30$-$360  \\
CDHSW       & Fe     & 20$-$212  \\
CHORUS      & Pb     & 10$-$200  \\
NuTeV       & Fe     & 30$-$500  \\
\hline
\end{tabular}
\end{center}
\normalsize
\end{wraptable}
%%%%%%%%%%%%%%%%%%%%%%%%%%%%%%%% table %%%%%%%%%%%%%%%%%%%%%%%%%%%%%%%%%%%%%%

Recent neutrino DIS experiments are listed in Table 1
\cite{nu-exp-98, chorus-06, nutev06}.
These  measurements have been done for heavy nuclear targets
of iron and lead. In order to extract the structure functions
of the nucleon, one needs to remove nuclear medium effects,
which may be assumed to be the same as the ones for the charged-lepton
scattering. However, nuclear modifications are generally 
different from the ones in the charged-lepton DIS.
We discuss these nuclear medium effects in this article.
In the near future, the MINER$\nu$A collaboration will measure
the nuclear structure functions for the targets,
helium, carbon, iron, and lead \cite{minerva}.
Their measurements should provide valuable information for
clarifying the nuclear effects in neutrino reactions.

In Fig. 1, data are shown for the $Q^2$ dependence of
the structure functions $F_2$ and $F_3$ \cite{nutev06}
of NuTeV, CCFR, and CDHSW measurements together with
NuTeV fit curves. We notice that the minimum $x$ is about
0.01 for $Q^2 >$1 GeV$^2$ due to fixed target measurements
in comparison with HERA collider data of $x_{min} \sim 10^{-4}$.
Therefore, the current neutrino data are valuable for determining
the PDFs at medium and large $x$ ($0.01<x<0.8$).
As shown in Eq. (\ref{eqn:f23}), neutrino reactions provide
information on quark flavor separations which are different
from the charged-lepton ones.

\vspace{-0.2cm}
%%%%%%%%%%%%%%%%%%%%%%%%%%%%%%%% figure %%%%%%%%%%%%%%%%%%%%%%%%%%%%%%%%%%%%%%
\noindent
\begin{figure}[h!]
\hspace{0.6cm}
\parbox[t]{0.46\textwidth}{
       \epsfig{file=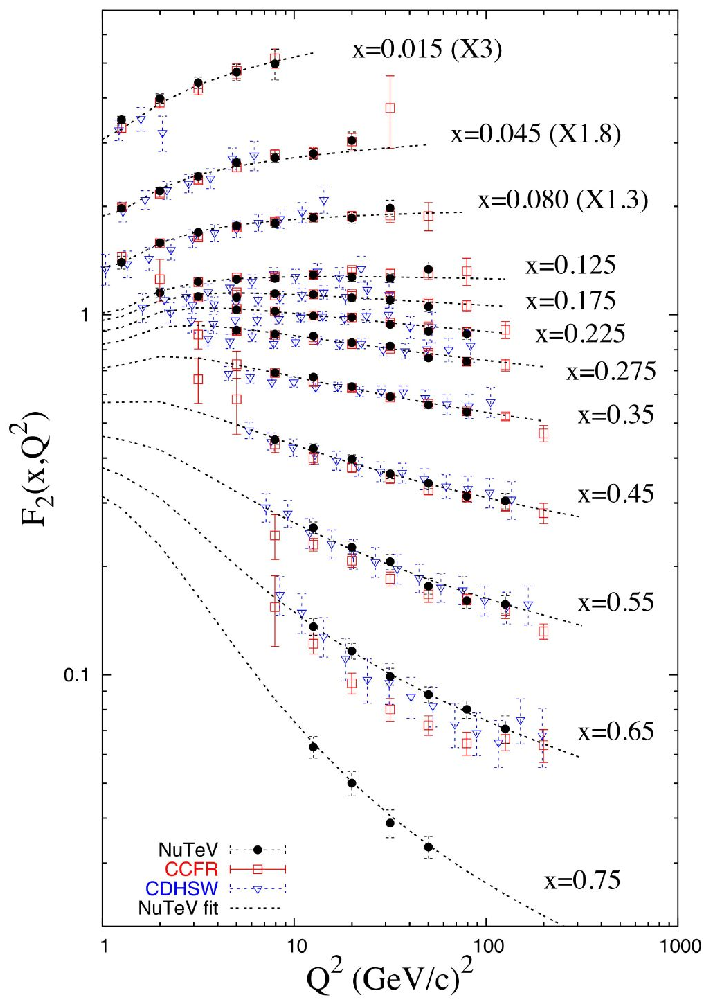,width=6.2cm}
}\hfill
\parbox[t]{0.46\textwidth}{
       \epsfig{file=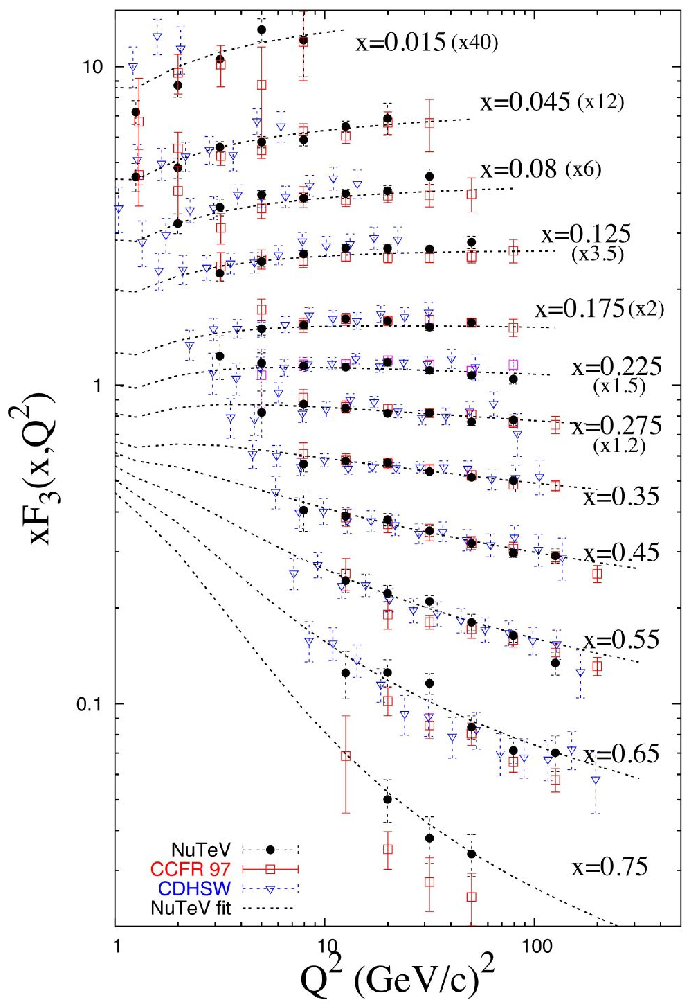,width=6.2cm} 
}
\vspace{-0.5cm}
\caption{Structure functions $F_2$ and $F_3$ 
         in neutrino deep inelastic scattering \cite{nutev06}.}
\label{fig:nu-f2-f3}
\end{figure}
%%%%%%%%%%%%%%%%%%%%%%%%%%%%%%%% figure %%%%%%%%%%%%%%%%%%%%%%%%%%%%%%%%%%%%%%

\vspace{-0.0cm}
%%%%%%%%%%%%%%%%%%%%%%%%%%%%%%%%%%%%%%%%%%%%%%%%%%%%%%%%%%%%%%%%%%%%%%%%%%%%%%%%
%%%%%%%%%%%%%%%%%%%%%%%%%%%%%%%%%%%%%%%%%%%%%%%%%%%%%%%%%%%%%%%%%%%%%%%%%%%%%%%%
\section{Determination of nuclear PDFs}
\vspace{-0.1cm}

%%%%%%%%%%%%%%%%%%%%%%%%%%%%%%%%%%%%%%%%%%%%%%%%%%%%%%%%%%%%%%%%%%%%%%%%%%%%%%%%
\subsection{Nucleonic PDFs and nuclear modifications}
\vspace{-0.2cm}

%%%%%%%%%%%%%%%%%%%%%%%%%%%% figure %%%%%%%%%%%%%%%%%%%%%%%%%%%%
\begin{wrapfigure}{r}{0.48\textwidth}
   \vspace{-0.25cm}
   \begin{center}
       \epsfig{file=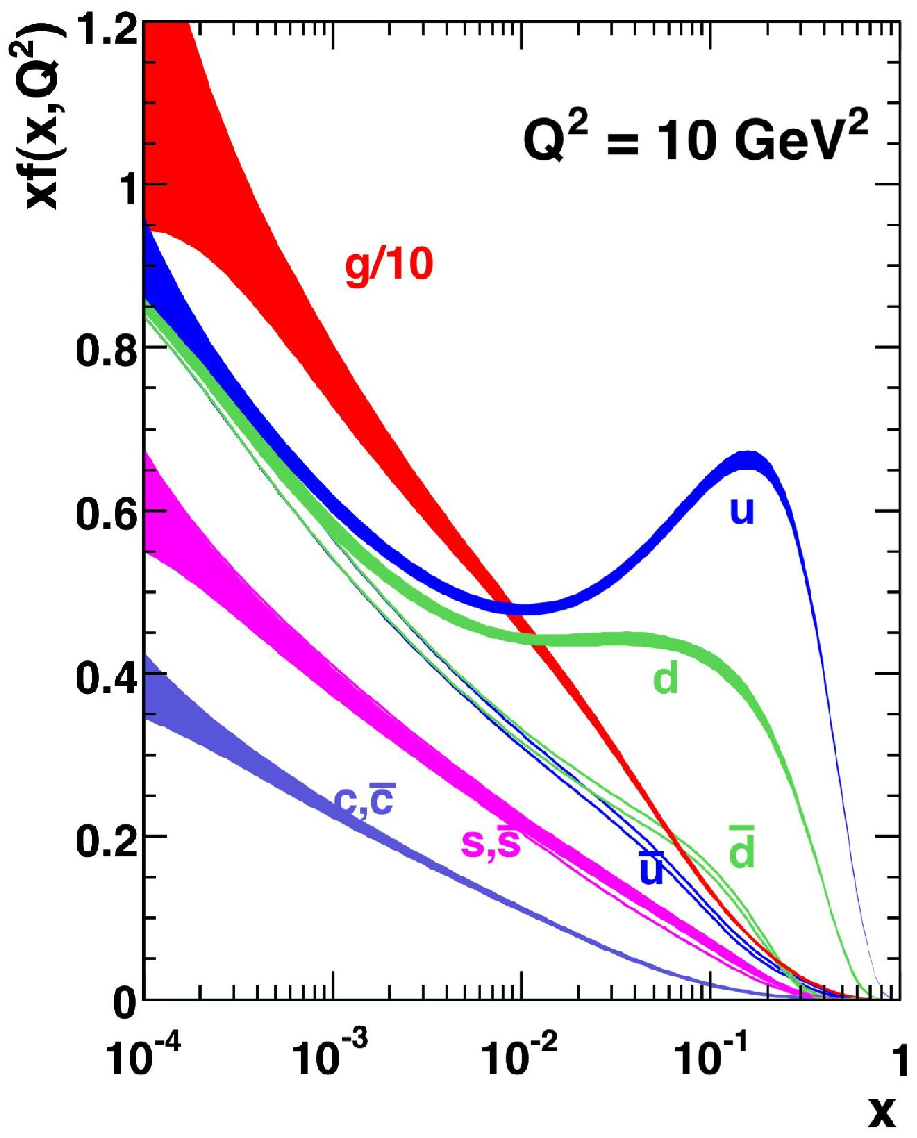,width=0.40\textwidth} \\
   \end{center}
   \vspace{-0.3cm}
       \begin{minipage}[c]{0.7cm}
       \ \ 
       \end{minipage}
       \begin{minipage}[c]{6.0cm}
       \setlength{\baselineskip}{10pt} 
       {\small {\bf FIGURE 2}: PDFs in the nucleon \cite{mstw09}}
       \end{minipage}
   \vspace{-0.2cm}
\label{fig:j-parc}
\end{wrapfigure}
%%%%%%%%%%%%%%%%%%%%%%%%%%%% figure %%%%%%%%%%%%%%%%%%%%%%%%%%%%

The structure functions $F_2$ and $F_3$ are expressed in terms of the PDFs,
which are then convoluted with coefficient functions for taking into
account higher-order effects of $\alpha_s$.
The PDFs of the nucleon have been investigated for a long time.
Because of a variety of experimental measurements such as charged-lepton
DIS, neutrino DIS, Drell-Yan, $W$ and jet production processes, etc.
and also established theoretical analysis techniques, 
the PDFs are now precisely determined from small to relatively
large $x$. A typical situation is shown in Fig. 2 at $Q^2$=10 GeV$^2$.
%%%
Uncertainties of the PDFs are shown by the bands in the figure.
They indicate that the distributions are well determined in the wide
range of $x$. Relative uncertainties ($\delta f/f$) are large
at large $x$ ($>0.5$) although they are not conspicuous in Fig. 2 
because the distributions themselves are small.

%%%%%%%%%%%%%%%%%%%%%%%%%%%% figure %%%%%%%%%%%%%%%%%%%%%%%%%%%%
\begin{wrapfigure}{r}{0.48\textwidth}
   \vspace{-0.25cm}
   \begin{center}
       \epsfig{file=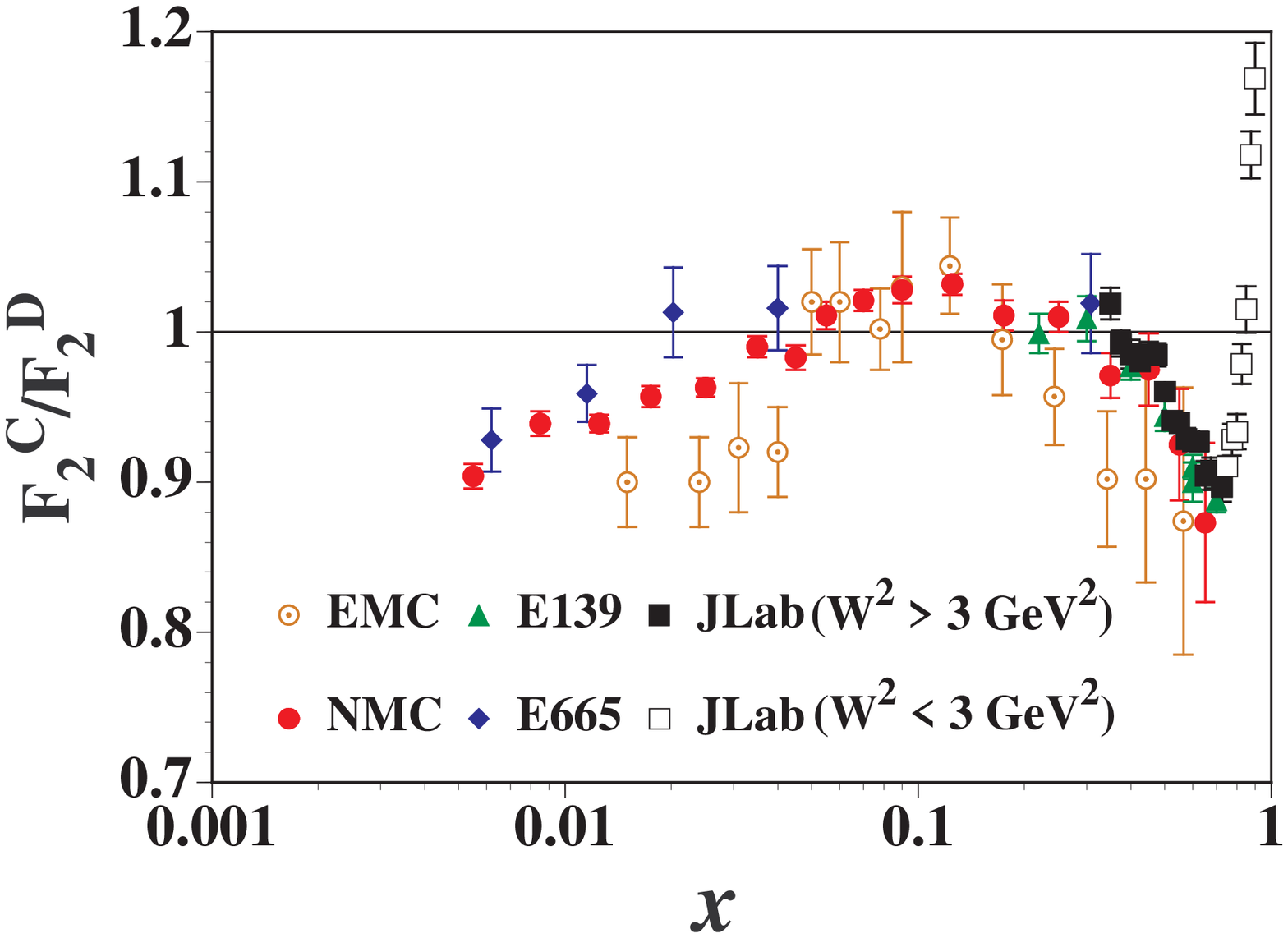,width=0.40\textwidth} \\
   \end{center}
   \vspace{-0.3cm}
       \begin{minipage}[c]{0.4cm}
       \ \ 
       \end{minipage}
       \begin{minipage}[c]{6.3cm}
       \setlength{\baselineskip}{10pt} 
       {\small {\bf FIGURE 3}: Nuclear modifications of $F_2$}
       \end{minipage}
   \vspace{-0.2cm}
\label{fig:j-parc}
\end{wrapfigure}
%%%%%%%%%%%%%%%%%%%%%%%%%%%% figure %%%%%%%%%%%%%%%%%%%%%%%%%%%%

Typical measurements on nuclear modifications of $F_2$ 
in charged-lepton scattering are shown
for the carbon nucleus as an example in Fig. 3, where
the ratio $F_2^C/F_2^D$ is shown \cite{hkn07,jlab-09}.
Here, $C$ and $D$ indicate the carbon nucleus and
the deuteron. The JLab data at large $x$ ($>0.8$) 
have small invariant masses ($W^2<$3 GeV$^2$) \cite{jlab-09}, 
so that they are not considered to be obtained in the DIS process.
Physics mechanisms for these nuclear medium effects are explained
in Ref. \cite{sumemc}. At large $x$ ($>0.7$), the ratio tends to
increase with $x$, which is caused by nucleon's Fermi motion in
a nucleus. At medium $x$ ($0.2<x<0.6$), main effects come from
nuclear binding possibly together with internal
nucleon modifications in a nuclear medium \cite{st94}.
The Fermi-motion and binding effects are theoretically calculated
in a convolution model. The nuclear structure function $F_2^A$ is
given by the free nucleonic one $F_2^N$ convoluted with a nucleon 
energy-momentum distribution, which is called a spectral function,
in a nuclear medium.
At small $x$ ($x<0.1$), the suppression of the ratio is caused
by shadowing, which is described by multiple scattering
of $q \bar q$-like configuration of the virtual photon.
The enhancement at $x \sim 0.15$ is called antishadowing.
It should be caused by the conservations of baryon number and
charge of a nucleus, but there is an attempt to attribute
it to a constructive interference effect in Pomeron and Reggeon exchanges
\cite{antishadow}.

%%%%%%%%%%%%%%%%%%%%%%%%%%%%%%%%%%%%%%%%%%%%%%%%%%%%%%%%%%%%%%%%%%%%%%%%%%%%%%%%
\subsection{Analysis method for nuclear PDFs}
\vspace{-0.2cm}

We explain global-analysis methods for determining nuclear PDFs.
A nucleus consists of mainly protons and neutrons although
the existence of mesons needs to be considered. 
Therefore, it is appropriate to assume that a NPDF (nuclear PDFs)
$f_i^A (x)$ for the parton type $i$ is given by proton and neutron
contributions as the first approximation, and then their nuclear
modifications are addressed, for example, 
in the following functional form \cite{hkm01,hkn04}:
\begin{equation}
f_i^A (x,Q_0^2) = w_i (x,A,Z) \, 
    \frac{1}{A} \left[ Z\,f_i^{p} (x,Q_0^2) 
                + (A - Z) f_i^{n} (x,Q_0^2) \right] ,
\end{equation}
where $p$ and $n$ indicate the proton and neutron,
and $Q_0^2$ is the initial $Q^2$ scale.
The isospin symmetry ($u \equiv d^n  = u^p ,\;d \equiv u^n  = d^p$)
is assumed in relating the distributions in the neutron to
the ones in the proton. The $x$-dependent functional form is not unique.
In the following, some recent parametrizations are shown as examples:
\begin{align}
& \bullet {\rm \ DS04 \, \text{\cite{ds04}} \ } (Q_0^2=0.4 \ {\rm GeV}^2 ): \
         f_i^{N/A} (x) = \int \,\frac{{dy}}{y}W_i (y,A,Z)\,
         f_i^N (x/y)  , 
         \ \ \ \ \ \ \ \ \ \ \ \ \ \ \ \ \ \ \ \ \ \ 
\nonumber \\
& \ \ \ \ \ 
W_i (y,A,Z) = 
\left\{
\begin{array}{ll}
A \, \left[
a_v \,\delta (1 - \varepsilon _v - y) 
        + (1 - a_v )\, \delta (1 - \varepsilon _{v'}  - y) 
    \right ]       & \ \\
        \ \ \ \ \ \ \ 
        + n_v \left( {\frac{y}{A}} \right)^{\alpha _v } 
              \left( {1 - \frac{y}{A}} \right)^{\beta _v }  
        + n_s \left( {\frac{y}{A}} \right)^{\alpha _s } 
              \left( {1 - \frac{y}{A}} \right)^{\beta _s } 
        & (i=V), \\
A\,\delta (1 - y) + \frac{a_{i}}{N_{i}}
          \left( \frac{y}{A} \right)^{\alpha _{i} } 
          \left( 1 - \frac{y}{A} \right)^{\beta _{i} } 
        & (i=s, \ g)
\end{array}
\right.
\\
& \bullet {\rm \ HKN07 \, \text{\cite{hkn07}} \ } (Q_0^2=1 \ {\rm GeV}^2 ): \  
         f_i^A (x) = w_i (x,A,Z)\,\frac{1}{A}
         \left[ {Z\,f_a^p (x) + (A - Z)f_a^n (x)} \right],
\nonumber \\
& \ \ \ \ \ 
w_i (x,A,Z) = 1 + \left( {1 - \frac{1}{{A^{\alpha} }}} \right)
\frac{{a_i  + b_i x + c_i x^2  + d_i x^3 }}{{(1 - x)^{\beta} }} ,
\\
& \bullet {\rm \ SYKMOO08 \, \text{\cite{sykmoo08}} \ } (Q_0^2 =1.69 \ {\rm GeV}^2 ): \  
         f_i^A (x) = \frac{1}{A}
         \left[ {Z\,f_a^{p/A} (x) + (A - Z)f_a^{n/A} (x)} \right],
\nonumber \\
& \ \ \ \ \ 
xf_i^{N/A} (x) = 
\left\{
\begin{array}{ll}
A_0 x^{A_1} (1 - x)^{A_2} e^{A_3 x} (1 + e^{A_4} x)^{A_5}  &
(i = u_v, \ d_v, \ g, \ \bar u + \bar d, \ \bar s)  \\
A_0 x^{A_1} (1 - x)^{A_2} + (1 + A_3 x)(1 - x)^{A_4}       &
(i = \bar d/\bar u), \\
\end{array}
\right.
\\
& \bullet {\rm \ EPS09 \, \text{\cite{eps09}} \ } (Q_0^2 =1.69 \  {\rm GeV}^2 ): \ 
         f_i^A (x) = R_i^A (x) \frac{1}{A}
         \left[ {Z\,f_a^{p} (x) + (A - Z)f_a^{n} (x)} \right],
\nonumber \\
& \ \ \ \ \ 
R_i^A (x) = 
\left\{
\begin{array}{ll}
a_0  + (a_1  + a_2 x)[e^{-x} - e^{-x_a}]  &
(x \le x_a : \rm{shadowing}) \\
b_0  + b_1 x + b_2 x^2  + b_3 x^3 &
(x_a  \le x \le x_e : \rm{\text{antishadowing}}). \\
c_0  + (c_1  - c_2 x)(1 - x)^{ - \beta } &
(x_e  \le x \le 1: \rm{EMC \ \& \ Fermi}) \\
\end{array}
\right.
\end{align}          

The parameters in these equations are determined 
by global $\chi^2$ analyses of world experimental data
on nuclear structure functions. Experimental data are 
generally obtained in different $Q^2$ points from $Q_0^2$.
The standard DGLAP evolution equations are used for evolving
the distributions to the experimental points.
There are three conditions to be satisfied for the NPDFs,
so that three parameters should be fixed by 
the following relations \cite{hkm01,hkn04}:
\begin{equation}
\begin{array}{ll}
{\text{Baryon number:}} & 
A \int dx \left[ \frac{1}{3} u_v^A (x) 
               + \frac{1}{3} d_v^A (x) \right] = A, \\ 
{\text{Charge:}} & 
A \int dx \left[ \frac{2}{3} u_v^A (x) 
               - \frac{1}{3} d_v^A (x) \right] = Z, \\ 
{\text{Momentum:}} & 
A \sum\limits_{i = q,\bar q,g} \int dx \, xf_i^A (x)  = A. \\ 
\end{array}
\end{equation}

%%%%%%%%%%%%%%%%%%%%%%%%%%%% figure %%%%%%%%%%%%%%%%%%%%%%%%%%%%
\begin{wrapfigure}{r}{0.46\textwidth}
   \vspace{-0.25cm}
   \begin{center}
       \epsfig{file=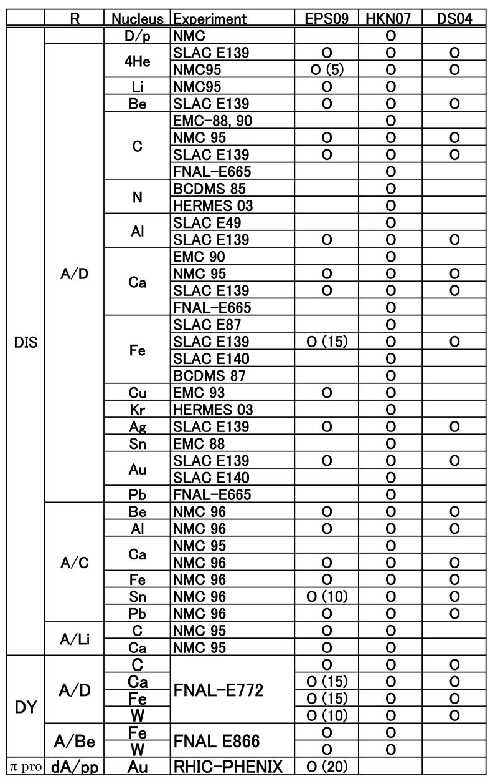,width=0.36\textwidth} \\
   \end{center}
   \vspace{-0.2cm}
       \begin{minipage}[c]{0.6cm}
       \ \ 
       \end{minipage}
       \begin{minipage}[c]{5.5cm}
       \setlength{\baselineskip}{10pt} 
       {\small {\bf TABLE 2}: Data for global analysis}
       \end{minipage}
   \vspace{0.2cm}
   \begin{center}
       \epsfig{file=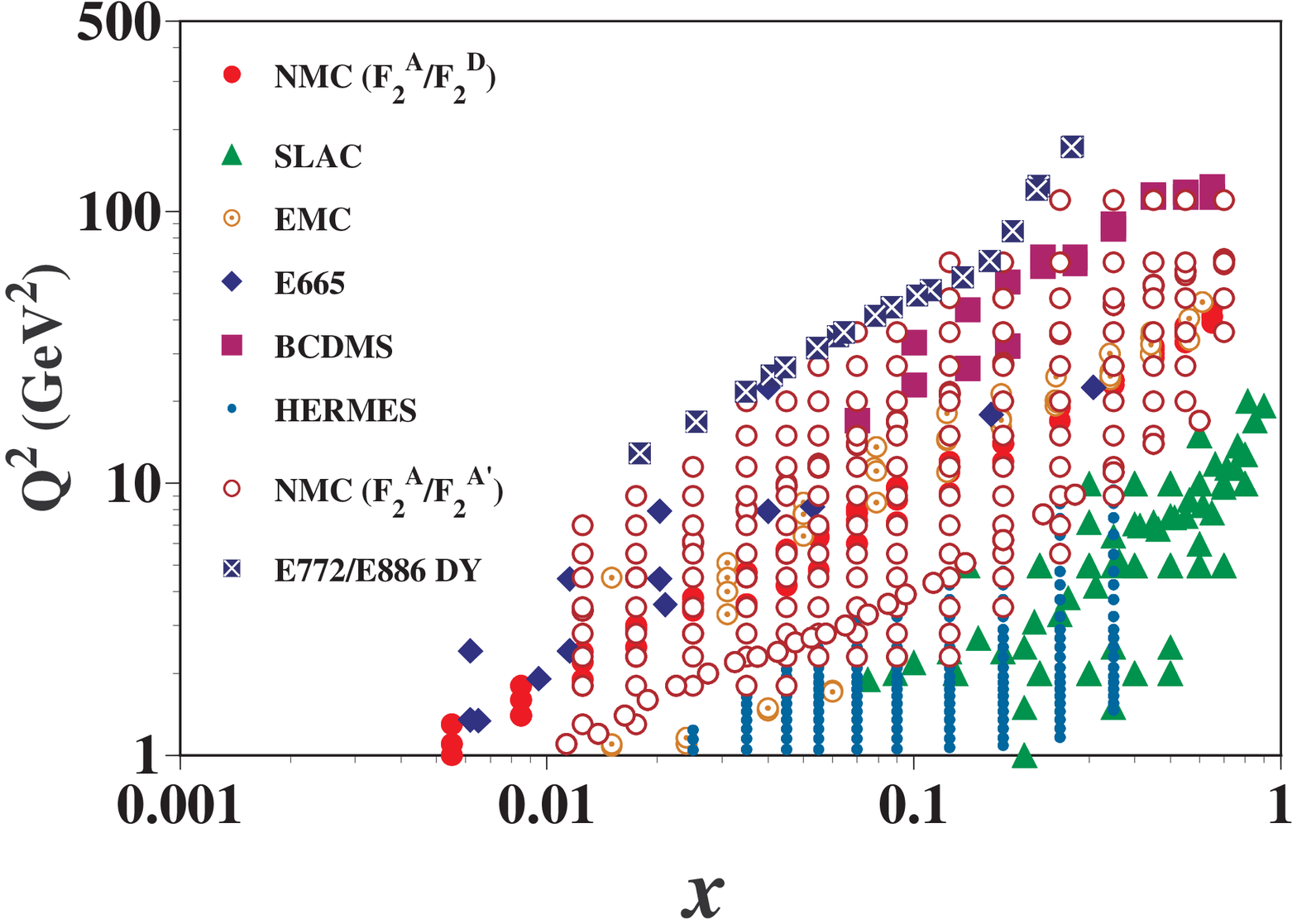,width=0.38\textwidth} \\
   \end{center}
   \vspace{-0.3cm}
       \begin{minipage}[c]{0.10cm}
       \ \ 
       \end{minipage}
       \begin{minipage}[c]{6.7cm}
       \setlength{\baselineskip}{10pt} 
       {\small {\bf FIGURE 4}: Kinematical range of data \cite{hkn04}}
       \end{minipage}
   \vspace{-0.2cm}
\end{wrapfigure}
%%%%%%%%%%%%%%%%%%%%%%%%%%%% figure %%%%%%%%%%%%%%%%%%%%%%%%%%%%

In the global analysis for the NPDFs, available data are still limited.
Furthermore, final-state interactions could affect cross sections
of hadron productions in nuclear reactions, and there are uncertainties
coming from fragmentation functions and their nuclear modifications.
The data used for the analyses are shown in Table 2.
There are measurements of the structure function ratios
$F_2^A/F_2^{A'}$, Drell-Yan cross section ratios 
$\sigma_{DY}^{pA}/\sigma_{DY}^{pA'}$, 
and pion-production ratios $\sigma_\pi^{dA}/\sigma_\pi^{pp}$ of RHIC.
The data sets used in the analyses, DS04, HKN07, and EPS09 are
shown in Table 2. The DS04 and EPS09 are somewhat selective
on the data sets. The numbers in the parentheses of EPS09 
indicate weight factors in calculating the total $\chi^2$. 
For example, the pion-production measurements of RHIC are
amplified with the weight factor 20.

The kinematical $(x,\, Q^2)$ range of $F_2$ and Drell-Yan data
is shown in Fig. 4. These data are taken by fixed-target experiments,
so that the range is limited, for example, in comparison
with the collider data of HERA. 
In future, the kinematical range will be extended by
Drell-Yan measurements at RHIC and LHC, and also
by the possible EIC (electron-ion collider) project.

A difference from the kinematical range $(0<x<1)$ of 
the nucleonic PDFs is that there exists a region $x>1$ for the NPDFs.
If the scaling variable is defined
by $x_A=Q^2 /(2 M_A \nu)$ with the target nuclear mass $M_A$ and
the energy transfer $\nu$, it is certainly restricted by 
$0<x_A<1$. However, experimental measurements are usually published
by the Bjorken scaling variable $x=Q^2 /(2 M_N \nu)$ even
for nuclei, so that the range of $x$ becomes $0<x<A$ because
of the mass ratio $M_A/M_N \approx A$.
However, the extremely large-$x$ region ($x>1$) is neglected
in current global analyses because 
there is no DIS data with $W^2 > 3$ GeV$^2$ as shown in Fig. 3 and
the structure functions are very small in this region.

From the $\chi^2$ fit to these data, the optimum values and
their errors are determined for the parameters of the NPDFs
at $Q_0^2$. Uncertainties of the NPDFs are estimated from
the global analyses. Popular methods are the Hessian 
and the Lagrange-multiplier methods.

We comment on the Kulagin and Petti's analysis \cite{kp07}.
Their approach is quite different from the above ones
in the sense that they try to calculate the nuclear corrections
in conventional nuclear models as far as they can, and then
they try to attribute remaining factors to off-shell effects 
of bound nucleons for explaining the data.
This off-shell part is parametrized and is determined
by analysis of nuclear structure functions.

\vfill\eject

%%%%%%%%%%%%%%%%%%%%%%%%%%%%%%%%%%%%%%%%%%%%%%%%%%%%%%%%%%%%%%%%%%%%%%%%%%%%%%%%
%%%%%%%%%%%%%%%%%%%%%%%%%%%%%%%%%%%%%%%%%%%%%%%%%%%%%%%%%%%%%%%%%%%%%%%%%%%%%%%%
\subsection{Nuclear PDFs and their uncertainties}
\vspace{-0.2cm}

%%%%%%%%%%%%%%%%%%%%%%%%%%%% figure %%%%%%%%%%%%%%%%%%%%%%%%%%%%
\begin{wrapfigure}{r}{0.32\textwidth}
   \vspace{-0.45cm}
   \begin{center}
       \epsfig{file=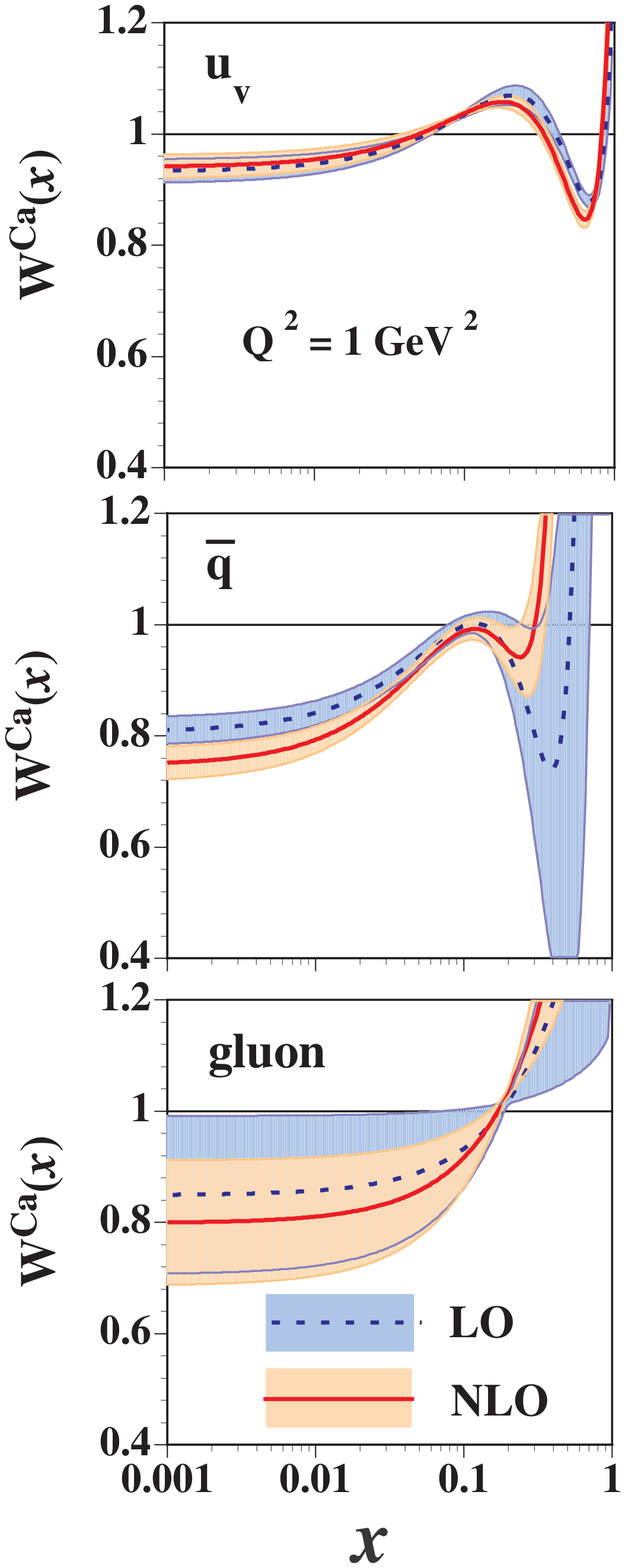,width=0.25\textwidth} \\
   \end{center}
   \vspace{-0.1cm}
       \begin{minipage}[c]{0.4cm}
       \ \ 
       \end{minipage}
       \begin{minipage}[c]{4.2cm}
       \setlength{\baselineskip}{10pt} 
       {\small {\bf FIGURE 5}: Modifications in calcium}
       \end{minipage}
   \vspace{-0.0cm}
\label{fig:j-parc}
\end{wrapfigure}
%%%%%%%%%%%%%%%%%%%%%%%%%%%% figure %%%%%%%%%%%%%%%%%%%%%%%%%%%%

Nuclear PDFs are determined by the data mainly on
the $F_2$ ratios and the Drell-Yan cross section ratios.
Total $\chi^2$/d.o.f. are 0.76 (DS04), 1.2 (HKN07), 
1.4 (SYKMOO08), and 0.80 (EPS09).
In the SYKMOO08 analysis, only the neutrino data are used.
Typical results are shown in Fig. 5 for the calcium nucleus.
Here, the nuclear modifications $w_i$ in Eq. (5) are shown 
for valence-quark, antiquark, and gluon distributions 
at $Q^2$=1 GeV$^2$ \cite{hkn07}.
Both leading order (LO) and next-to-leading order (NLO) results
are shown by the solid and dashed curves, respectively. Their
uncertainty ranges are shown by the shaded bands.

The figure indicates that the uncertainties are slightly
smaller in the NLO especially in antiquark and gluon
distributions. The results show that the valence-quark
distributions are well determined due to the accurate measurements
of the $F_2$ modifications at medium $x$ in the charged-lepton DIS.
The antishadowing of $F_2$ should be interpreted by 
the enhancement of the valence-quark distributions
at $x \sim 0.15$ because modifications are small for
antiquark distributions according to Fermilab Drell-Yan
measurements. The small-$x$ behavior of $w^A (x)$ for
the valence quarks is constrained by the baryon-number
and charge conservations.
%%%
The antiquark modifications are also well determined
at small $x$ ($<0.1$); however, there are large uncertainties
at $x>0.2$. This issue will be clarified in the near future
by the Fermilab E906 \cite{e906} and a possible J-PARC
Drell-Yan experiments \cite{j-parc}. The antiquark modifications
are still assumed to be flavor symmetric \cite{flavor}.
The gluon modifications are not determined well in the whole-$x$ range.
We need accurate measurements on direct-photon and jet productions
at RHIC and LHC.

%%%%%%%%%%%%%%%%%%%%%%%%%%%% figure %%%%%%%%%%%%%%%%%%%%%%%%%%%%
\begin{wrapfigure}{r}{0.55\textwidth}
   \vspace{-0.30cm}
   \begin{center}
       \epsfig{file=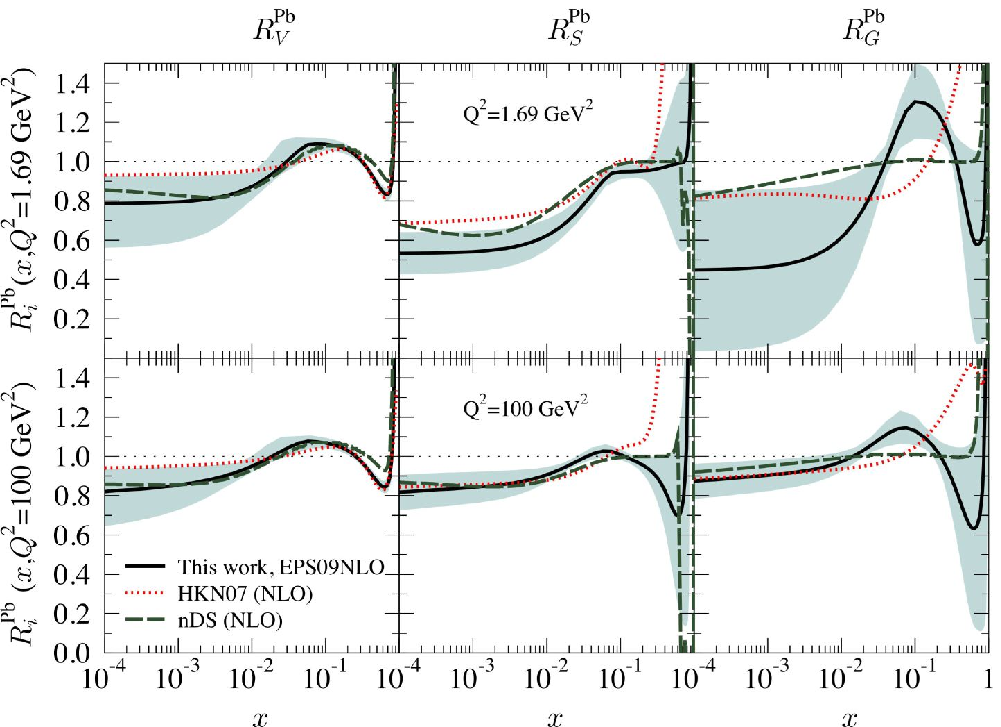,width=0.50\textwidth} \\
   \end{center}
   \vspace{-0.2cm}
       \begin{minipage}[c]{1.0cm}
       \ \ 
       \end{minipage}
       \begin{minipage}[c]{7.5cm}
       \setlength{\baselineskip}{10pt} 
       {\small {\bf FIGURE 6}: Modifications in lead \cite{eps09}}
       \end{minipage}
   \vspace{-0.3cm}
\label{fig:j-parc}
\end{wrapfigure}
%%%%%%%%%%%%%%%%%%%%%%%%%%%% figure %%%%%%%%%%%%%%%%%%%%%%%%%%%%

We show the results for the lead nucleus and comparison of various
NPDF results in Fig. 6 \cite{eps09}. Nuclear modifications are shown
for the valence-quark, sea-quark, and gluon distributions
($R_V^{Pb}$, $R_S^{Pb}$, $R_G^{Pb}$) at $Q^2$=1.69 GeV$^2$
and 100 GeV$^2$. All the analyses results (DS04, HKN07, EPS09)
are shown in the NLO.
The uncertainty regions of the EPS09 are indicated by the shaded bands.
All the three analyses use similar data sets as shown in Table 2.
The functional forms are slightly different in these models.
However, irrespective of these differences, the three modifications
are roughly within the error bands, which indicates consistency
of these results.

%%%%%%%%%%%%%%%%%%%%%%%%%%%% figure %%%%%%%%%%%%%%%%%%%%%%%%%%%%
\begin{wrapfigure}{r}{0.40\textwidth}
   \vspace{-0.60cm}
   \begin{center}
       \epsfig{file=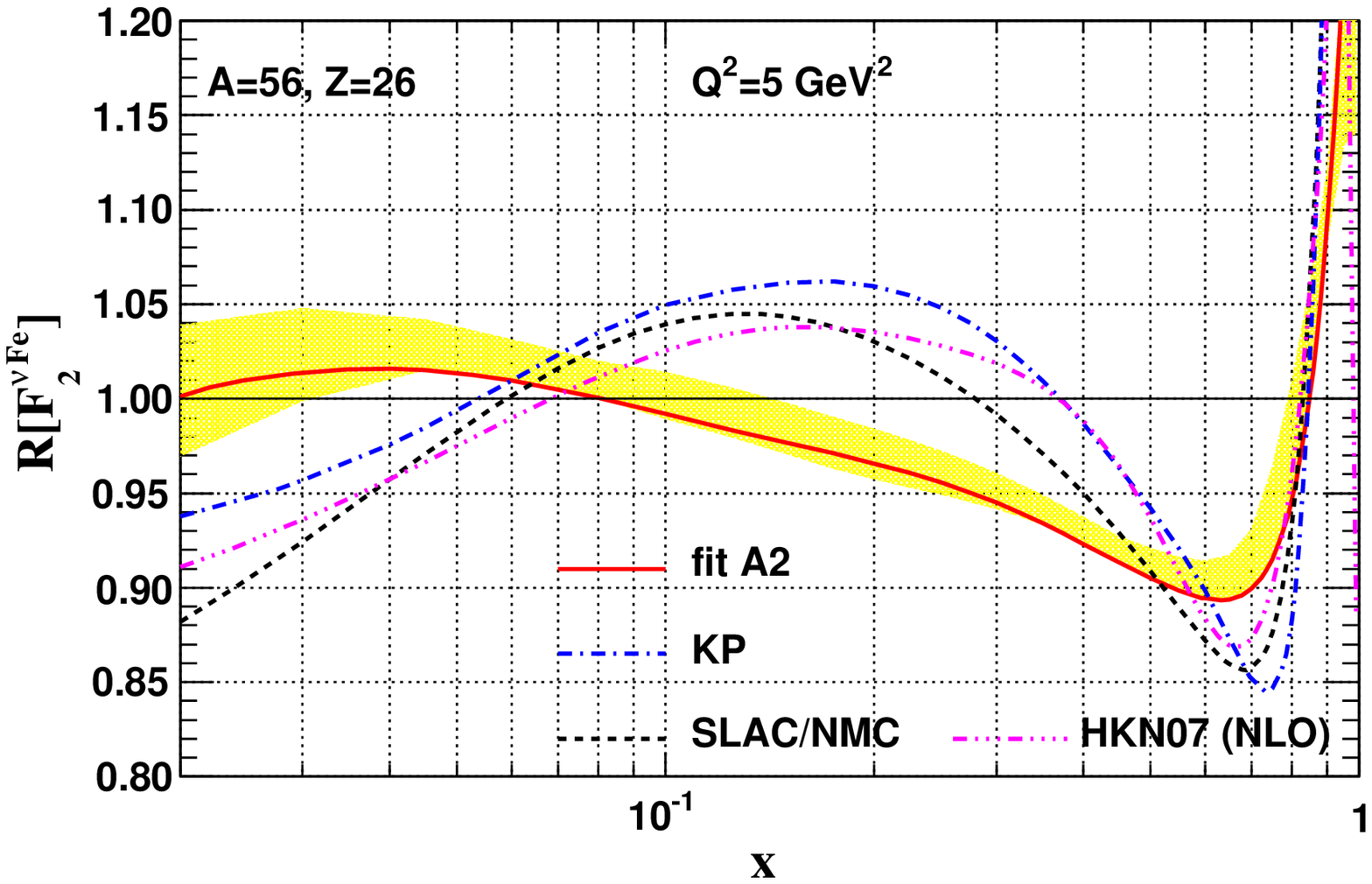,width=0.38\textwidth} \\
   \end{center}
   \vspace{-0.5cm}
       \begin{minipage}[c]{0.3cm}
       \ \ 
       \end{minipage}
       \begin{minipage}[c]{5.0cm}
       \setlength{\baselineskip}{10pt} 
       {\small {\bf FIGURE 7}: Modifications in charged-lepton and
       neutrino scattering off iron \cite{sykmoo08}}
       \end{minipage}
   \vspace{-0.2cm}
\label{fig:j-parc}
\end{wrapfigure}
%%%%%%%%%%%%%%%%%%%%%%%%%%%% figure %%%%%%%%%%%%%%%%%%%%%%%%%%%%

In the analyses of Figs. 5 and 6, neutrino data are not 
included. In Ref. \cite{sykmoo08}, a global analysis
was done for the neutrino data. A typical result is shown
as the ``fit $A2$" in Fig. 7 for the modifications of 
$F_2^{\nu Fe}$ in comparison with a charged-lepton
modification curve (SLAC/NMC) and other analysis results
(KP and HKN07). Obviously, there are unexpected differences
between the modifications of charged lepton and neutrino
reactions. The medium-$x$ depletion is smaller
in the neutrino scattering, and the antishadowing region 
is shifted toward a smaller-$x$ region.
Because the neutrino DIS data are corrected mostly by
assuming both nuclear modifications are the same, this 
discrepancy poses a question on the precision
of current nucleonic PDFs by using neutrino DIS data
for the nuclear targets.
A similar discrepancy was reported by the recent MSTW 
analysis \cite{mstw09}, so that it should be clarified
by future theoretical and experimental investigations.

\vspace{-0.3cm}
%%%%%%%%%%%%%%%%%%%%%%%%%%%%%%%%%%%%%%%%%%%%%%%%%%%%%%%%%%%%%%%%%%%%%%%%%%%%%%%%
%%%%%%%%%%%%%%%%%%%%%%%%%%%%%%%%%%%%%%%%%%%%%%%%%%%%%%%%%%%%%%%%%%%%%%%%%%%%%%%%
\begin{theacknowledgments}
\vspace{-0.2cm}
The authors would like to thank M. Sakuda for his support on this project. 
They thank D. Naples, C. A. Salgado, I. Schienbein, G. Watt, 
American Physical Society, 
European Physics Journal,
and Journal of High Energy Physics
for permitting them to use figures.
\end{theacknowledgments}

\vspace{-0.3cm}
%%%%%%%%%%%%%%%%%%%%%%%%%%%%%%%%%%%%%%%%%%%%%%%%%%%%%%%%%%%%%%%%%%%%%%%%%%%%%%%%
%%%%%%%%%%%%%%%%%%%%%%%%%%%%%%%%%%%%%%%%%%%%%%%%%%%%%%%%%%%%%%%%%%%%%%%%%%%%%%%%

%%%%%%%%%%%%%%%%%%%%%%%%%%%%%%%%%%%%%%%%%%%%%%%%%%%%%%%%%%%%%%%%%%%%%%%%%%%%%%%%
%%%%%%%%%%%%%%%%%%%%%%%%%%%%%%%%%%%%%%%%%%%%%%%%%%%%%%%%%%%%%%%%%%%%%%%%%%%%%%%%

\end{document}